\documentclass[twocolumn,english,showpacs,showkeys,reprint,linenumbers]{revtex4}
\usepackage[utf8]{inputenc}
\usepackage{array}
\usepackage{textcomp}
\usepackage{amsmath}
\usepackage{amssymb}
\usepackage{graphicx}
\usepackage{setspace}

\makeatletter

\providecommand{\tabularnewline}{\\}

\@ifundefined{textcolor}{}
{%
 \definecolor{BLACK}{gray}{0}
 \definecolor{WHITE}{gray}{1}
 \definecolor{RED}{rgb}{1,0,0}
 \definecolor{GREEN}{rgb}{0,1,0}
 \definecolor{BLUE}{rgb}{0,0,1}
 \definecolor{CYAN}{cmyk}{1,0,0,0}
 \definecolor{MAGENTA}{cmyk}{0,1,0,0}
 \definecolor{YELLOW}{cmyk}{0,0,1,0}
 }

\usepackage{babel}

\usepackage{babel}

\makeatother

\usepackage{babel}
\begin{document}

\title{Micro-strip ferromagnetic resonance study of strain-induced anisotropy
in amorphous FeCuNbSiB film on flexible substrate}

\author{Fatih Zighem$^{1}$}

\email{zighem@univ-paris13.fr}

\author{Anouar El Bahoui$^{1}$}

\author{Johan Moulin$^{2}$}

\author{Damien Faurie$^{1}$}

\email{faurie@univ-paris13.fr}

\author{Mohamed Belmeguenai$^{1}$}

\author{Silvana Mercone$^{1}$}

\author{Halim Haddadi$^{3}$}

\affiliation{$^{1}$Laboratoire des Sciences des Procédés et des Matériaux, CNRS-Université
Paris XIII, Sorbonne Paris Cité, Villetaneuse, France}

\affiliation{$^{2}$Institut d'Electronique Fondamentale, UMR 8622, Université
Paris Sud-CNRS, Orsay, France}

\affiliation{$^{3}$ Laboratoire MSMP - Carnot Arts, ENSAM ParisTech, rue Saint-Dominique,
51006 Châlons-en-Champagne}

\date{September 1$^{th}$ 2014 }
\begin{abstract}
The magnetic anisotropy of a FeCuNbSiB (Finemet\textregistered{})
film deposited on Kapton\textregistered{} has been studied by micro-strip
ferromagnetic resonance technique. We have shown that the flexibility
of the substrate allows a good transmission of elastic strains generated
by a piezoelectric actuator. Following the resonance field angular
dependence, we also demonstrate the possibility of controlling the
magnetic anisotropy of the film by applying relatively small voltages
to the actuator. Moreover, a suitable model taking into account the
effective elastic strains measured by digital image correlation and
the effective elastic coefficients measured by Brillouin light scattering,
allowed to deduce the magnetostrictive coefficient. This latter was
found to be positive $(\lambda=16\times10^{-6}$) and consistent with
the usually reported values for bulk amorphous FeCuNbSiB. 
\end{abstract}


\maketitle

\section{Introduction}

The strain control of magnetization orientation distribution in magnetic
thin films \textit{via} the magnetoelastic or the magnetostrictive
properties, is increasingly studied to face new challenges in magnetoelectronics
and spintronics \cite{Nan2011_Adv_Mat,Ramesh2010_Adv_Mat,Martins2013_Adv_Func_Mater,Roy2011,Ma2011,Lahtinen2011}.
An easy way for mastering the magnetization is to make adhering the
thin film on a piezoelectric actuator, so that the strains can be
applied to the film by varying continuously a voltage on the piezoelectric
actuator. This latter will succeed in transferring the strains in
a more or less efficient way, depending on the chosen system. Obviously,
more the elastic strains are transferred at the interface, more the
indirect magnetoelectric effect is optimized. Generally, at a first
step, the magnetic thin films are deposited on a substrate of usually
one hundred microns thick. At a second step the film/substrate is
cemented on an actuator \cite{Pettiford2008,Zighem_JAP2013,Brandlmaier2008_PRB,Tiercelin2011}.
At this point two interfaces will play their role in the transmission
of strains: the actuator/substrate interface and the substrate/film
one. This double transfer limits the desired phenomenon, especially
when the substrate is stiff such as for commonly used wafer (Si, GaAs,
...) \cite{Brandlmaier2008_PRB_Bis}. Besides, in this configuration,
it is generally hard to predict perfectly the amount of the induced
strains inside the thin film ($\varepsilon_{xx}$ and $\varepsilon_{yy}$)
even if the piezoelectric coefficients $d_{ijk}$ of the actuator
are known. This is due to the partial strains transmissions from the
actuator to the substrate, especially in the case of stiff substrates.
This well-known reported limitation (a few ten percents of losses
in best cases) can be avoided by depositing the magnetic thin film
on a compliant substrate such as polyimides that are more and more
used in flexible spintronics \cite{Barrault2012,Bedoya-Pinto2014,ZhangJAP2013_Bis}.
More precisely we reported in Ref. \cite{Zighem_JAP2013}, by comparing
film and actuator strains, that one advantage of studying thin films
on polymer substrates is the very good strains transmissions (nearly
100\%) resulting from the substrate compliance. However, once well-known
strains are transferred to the film, the estimation of the stresses
($\sigma_{xx}$ and $\sigma_{yy}$) is straightforward only if the
elastic coefficients of this latter are known. Otherwise Hook's law
is not applicable. Unfortunately, the elastic coefficients are not
always well-known, this is especially not the case when studying thin
film of new functional alloys.

\begin{figure}
\includegraphics[bb=20bp 220bp 610bp 585bp,clip,width=7.5cm]{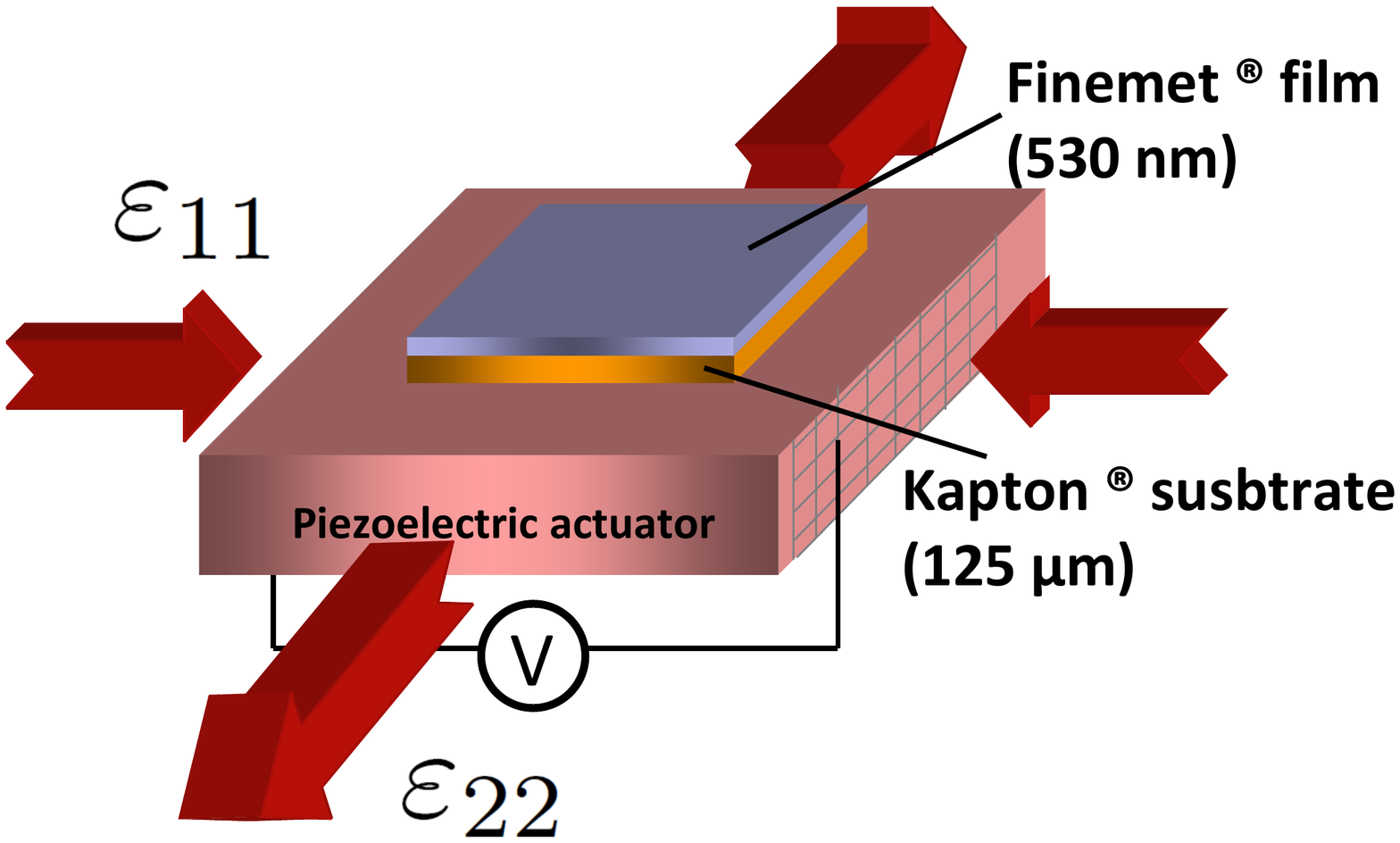}

\caption{Sketch of the studied heterostructure showing the 530 nm thick Finemet\textregistered{}
film deposited onto a 125 $\text{\textmu}$m thick Kapton\textregistered{}
substrate and glued onto the piezoelectric actuator. The arrows qualitatively
represent the in-plane strains of the piezoelectric actuator.}

\label{Fig_Sketch_Actuator} 
\end{figure}

In this paper, voltage induced strain effect on the in-plane magnetic
anisotropy has been quantitatively studied in a ferromagnetic film
deposited on a flexible substrate and glued onto a piezoelectric actuator.
In particular, the structural (Young's modulus $E$ and Poisson's
ratio $\nu$) and magnetoelastic (magnetostriction coefficient at
saturation $\lambda$) properties have been completely determined
by combining different techniques. Our general methodology is based
on the combination from one hand of i) the control of the applied
stress state in a magnetic thin film and from another hand on ii)
the magnetic uniform precession mode resonance field measurement as
function of the stress. In order to control the stress state in the
magnetic film, the film/substrate system is glued onto a piezoelectric
actuator that allows applying in-plane strains ($\varepsilon_{11}$
and $\varepsilon_{22}$) when applying voltage to the actuator as
it is shown on Figure \ref{Fig_Sketch_Actuator}. To properly estimate
the in-plane strains by the double interfaces, we adopted the Digital
Image Correlation (DIC) technique based on the optical observation
of the film surface and its evolution during the film straining (section
IV).

A non-destructive method (Brillouin light scattering \cite{Rossignol2004,Fillon2014,Pham2013})
has been used to quantify the elastic coefficients of the amorphous
magnetic thin film (BLS study in section III). The magnetic anisotropy
dependence upon the applied stress has been followed by measuring
the behavior of the high frequency resonance field while applying
an electric field inside the actuator. This resonance field is directly
linked to the magnetic anisotropy and it can be estimated by performing
Micro-Strip FerroMagnetic Resonance (MS-FMR )\cite{belmeguenai2009,belmeguenai2013}
experiments in several geometrical configurations (section IV). Moreover,
by adjusting an appropriate model (section II) to the experimental
data, we have shown how to estimate with accuracy the thin film effective
magnetostriction coefficient at saturation ($\lambda$).

\section{theoretical background}

\begin{figure}
\includegraphics[bb=150bp 230bp 660bp 580bp,clip,width=8.5cm]{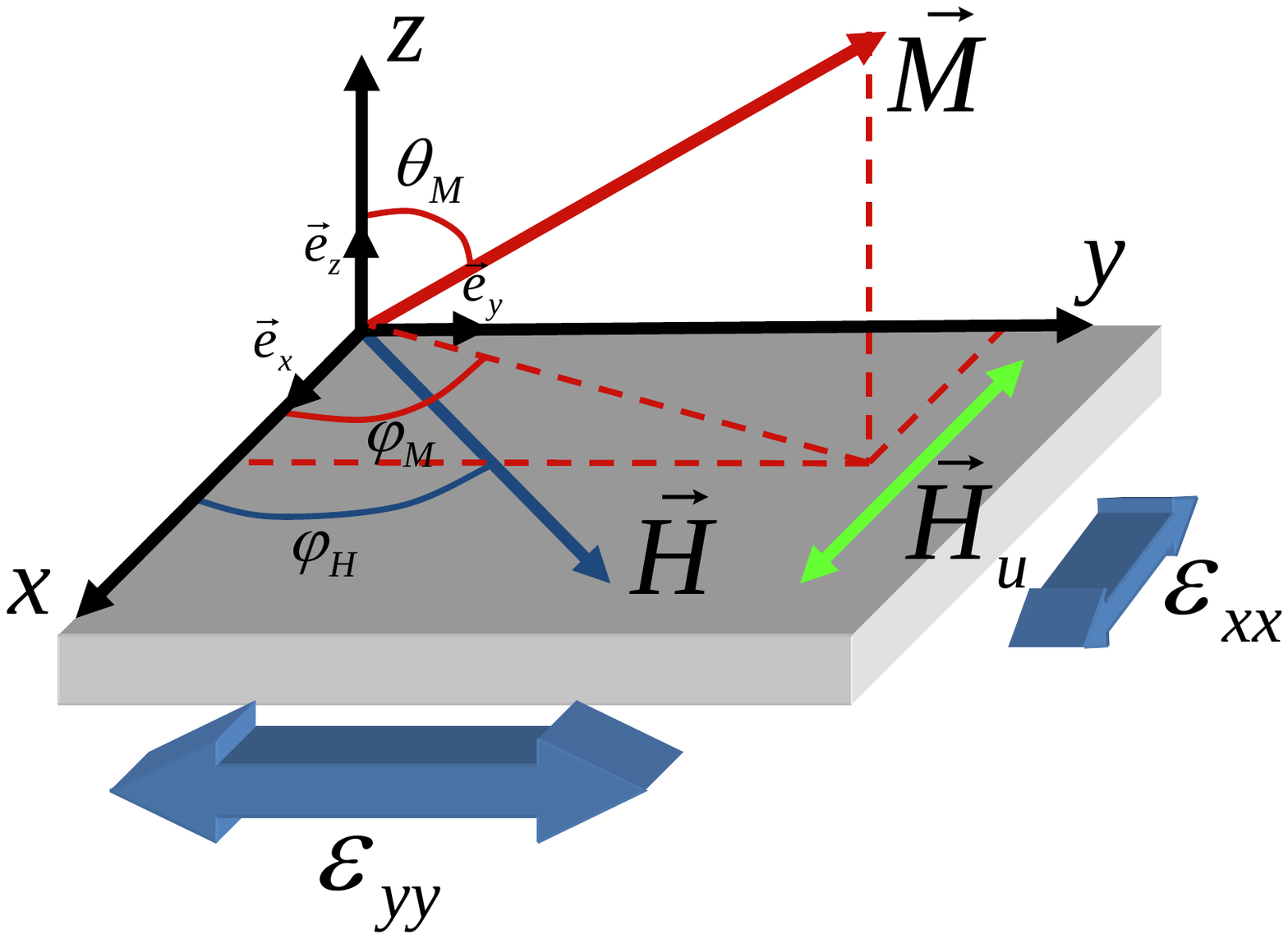}

\label{Fig_Sketch_FMR}

\caption{Schematic illustration showing angles, fields and coordinate systems
used in the text. }
\end{figure}

The voltage-induced strain effect has been experimentally studied
by ferromagnetic resonance through uniform precession mode resonance
field as a function of the applied voltage. Indeed, the resonance
field (or resonance frequency) of the uniform precession mode is influenced
by the magnetoelastic behavior of the thin film. All the experimental
spectra presented in this work have been performed at room temperature.
Moreover, they have been done at relatively {}``high'' applied magnetic
field in order to have a uniform magnetization inside the film. The
expression of the resonance field is derived taking into account in-plane
strains ($\varepsilon_{xx}$ and $\varepsilon_{yy}$). These in-plane
strains will be induced by applying an electric field inside the piezoelectric
actuator. When any magnetoelastic contribution is required, the magnetic
energy density of a ferromagnetic thin film, using the coordinates
system of \ref{Fig_Sketch_FMR}, can be written as:

\begin{equation}
F_{0}=F_{zee}+F_{dip}+F_{exch}+F_{anis}
\end{equation}

Where the three first terms stand for the Zeeman, the dipolar and
the exchange contributions, respectively. The last term, which corresponds
to the anisotropy contribution, will be written as an \textit{ad hoc}
in-plane uniaxial anisotropy characterized by the anisotropy constant
$K_{u}$. This is correct when no out-of-plane or surface contributions
to the anisotropy are observed in the film. This term can then be
written as: 
\begin{equation}
F_{anis}=-\frac{K_{u}}{M_{s}^{2}}\left(\vec{M}\cdot\vec{u}_{x}\right)^{2}
\end{equation}

$\vec{u}_{x}$ is a unit vector along the {}``easy axis'' (along
$x$ direction) while $\vec{M}$ and $M_{s}$ are respectively the
vector and the module of the magnetization. The stress effect will
be modeled through a magnetoelastic density of energy $F_{me}$ which
will be added to $F_{0}$: 
\begin{equation}
F_{me}=-\frac{3}{2}\lambda\left(\left(\gamma_{x}^{2}-\frac{1}{3}\right)\sigma_{xx}+\left(\gamma_{y}^{2}-\frac{1}{3}\right)\sigma_{yy}\right)
\end{equation}

$\sigma_{xx}$ and $\sigma_{yy}$ being the in-plane principal stress
tensor components while $\gamma_{x}$ and $\gamma_{y}$ correspond
to the direction cosines of the in-plane magnetization. The presence
of a unique saturation magnetostriction coefficient $\lambda$ is
due to the amorphous structure of the thin film. The relation between
the principal stress components ($\sigma_{xx}$, $\sigma_{yy}$ )
and strains ($\varepsilon_{xx}$, $\varepsilon_{yy}$) tensors is
thus given by an isotropic Hook's law where $E$ is the Young's modulus
and $\nu$ is the Poisson's ratio:

\begin{flalign}
\sigma_{xx} & =\left(\frac{E}{1+\nu}\right)\left(\frac{1}{1-\nu}\varepsilon_{xx}+\frac{\nu}{1-\nu}\varepsilon_{yy}\right)\label{eq:Hook_law_1}\\
\sigma_{yy} & =\left(\frac{E}{1+\nu}\right)\left(\frac{1}{1-\nu}\varepsilon_{yy}+\frac{\nu}{1-\nu}\varepsilon_{xx}\right)\label{eq:Hook_law_2}
\end{flalign}
In these conditions, minimizing the total volume magnetic energy density
(i. e. $F=F_{0}+F_{me}$), the resonance field of the uniform precession
mode can be obtained thanks to the following relation \cite{Smit_Beljers_1955,Suhl_1955}:
\begin{equation}
\left(\frac{2\pi f}{\gamma}\right)^{2}=\left(\frac{1}{\sin\theta_{M}}\right)^{2}\left(\frac{\partial^{2}F}{\partial\theta_{M}^{2}}\frac{\partial^{2}F}{\partial\varphi_{M}^{2}}-\left(\frac{\partial^{2}F}{\partial\theta_{M}\varphi_{M}}\right)^{2}\right)\label{eq:Smit_Beljers}
\end{equation}

Where $f$ is the microwave driving frequency. The different energy
derivatives are calculated at the equilibrium direction of the magnetization.
In the above expression, $\gamma$ is the gyromagnetic factor $\gamma=g\times8.794\times10^{6}$
s$^{-1}$.Oe$^{-1}$ while $\theta_{M}$ and $\varphi_{M}$ stand
for the polar and the azimuthal angles of the magnetization. Note
that for an in-plane applied magnetic field, the equilibrium polar
angle is $\varphi_{M}=\frac{\pi}{2}$ because of the large effective
demagnetizing field associated with the planar film geometry and an
explicit expression is obtained for $f$: 
\begin{equation}
f^{2}=\left(\frac{\gamma}{2\pi}\right)^{2}H_{1}H_{2}
\end{equation}

where:

\begin{widetext}

\begin{flalign}
H_{1} & =4\pi M_{s}+H_{res}\cos\left(\varphi-\varphi_{H}\right)+\frac{2K_{u}}{M_{s}}\cos^{2}\varphi+\frac{3\lambda}{M_{s}}\left(\sigma_{xx}\cos^{2}\varphi+\sigma_{yy}\sin^{2}\varphi\right)\\
H_{2} & =\left(\frac{2K_{u}}{M_{s}}+\frac{3\lambda}{M_{s}}\left(\sigma_{xx}-\sigma_{yy}\right)\right)\cos2\varphi+H_{res}\cos\left(\varphi-\varphi_{H}\right)
\end{flalign}

\end{widetext}

Here $\varphi_{H}$ is the angle between the in-plane applied magnetic
field and the magnetic easy axis ($x$ direction). The analysis can
be simplified if the resonance field is larger than the effective
uniaxial anisotropy and magnetoelastic field: $\vec{H}_{u}=-\vec{\nabla}_{\vec{M}}F_{anis}$
and $\vec{H}_{me}=-\vec{\nabla}_{\vec{M}}F_{me}$, respectively. Indeed,
in this condition, the magnetization direction will be almost parallel
to the applied magnetic field ($\varphi_{M}\sim\varphi_{H}$). The
resonance field is thus given by:

\begin{widetext}

\begin{multline}
H_{res}=\frac{1}{2}\left(\sqrt{\left(4\pi M_{s}+\left(H_{u}+\frac{3\lambda}{M_{s}}\sigma_{xx}\right)\sin^{2}\varphi+\frac{3\lambda}{M_{s}}\sigma_{yy}\cos^{2}\varphi_{H}\right)^{2}+4\left(\frac{2\pi f}{\gamma}\right)^{2}}-4\pi M_{s}\right)\\
-H_{u}\left(\frac{1}{4}+\frac{3}{4}\cos2\varphi_{H}\right)-\frac{3\lambda}{M_{s}}\sigma_{xx}\left(\frac{1}{4}+\frac{3}{4}\cos2\varphi_{H}\right)-\frac{3\lambda}{M_{s}}\sigma_{yy}\left(\frac{1}{4}-\frac{3}{4}\cos2\varphi_{H}\right)\label{eq:Resonance_field}
\end{multline}

\end{widetext}

The first term essentially represents a constant shift in the resonance
field baseline because $4\pi M_{s}$ and $\frac{2\pi f}{\gamma}$
are found to be larger than the magnetoelastic and the uniaxial anisotropy
fields. The other terms correspond to the angular variation of the
resonance field due to the uniaxial anisotropy field (second term)
and to the voltage induced magnetoelastic anisotropy field (third
and fourth terms).

\section{As-deposited thin film characterization at zero stress applied}

\subsubsection{Finemet thin film deposition and structure}

An amorphous 530 nm thick Finemet\textregistered{} film was deposited
onto a 125 $\mu$m thick polyimide flexible substrate (Kapton\textregistered{})
by radio frequency sputtering. The deposition residual pressure was
of around 10$^{-7}$ mbar, while a working Ar pressure was of 40 mbar
and the RF power was of 250 W. A 10 nm thick Ti buffer layer was deposited
on the substrate to ensure a proper adhesion of the Finemet\textregistered{}
film. Finally, another 10 nm thick Ti cap layer was deposited on the
top of the Finemet\textregistered{} film in order to protect it from
oxidation. The composition of the film has been measured by EDS (Energy
Dispersive Spectroscopy) and is close to that of the target (Fe$_{73.5}$Cu$_{1}$Nb$_{3}$Si$_{15.5}$B$_{7}$)
while the thickness of the film (530 nm) has been measured by Scanning
Electron Microscopy and mechanical profilometry.

The film/substrate system has been then glued onto a piezoelectric
actuator. Figure \ref{Fig_Sketch_Actuator} presents a sketch of the
studied heterostructure for studying the so-called inverse magnetoelectric
effect. In this Figure $\varepsilon_{11}$ and $\varepsilon_{22}$
correspond to the in-plane strains at the surface of the actuator
(top) and at the interface with the flexible substrate. This latter
is again used here as it is the best possible medium for efficiently
transfer the in-plane strain between the actuator and the ferromagnetic
film. Maximum values of in-plane strains at a fixed voltage will be
obtained as compared to those obtained when using a rigid substrate.
Indeed, there is roughly two orders of magnitude between the Young's
modulus values of rigid substrates and the ones of a flexible one
($\sim$4 GPa for Kapton\textregistered{}) and $\sim$180 GPa for
Si). Nevertheless, even if the interacting vector is the voltage induced
in-plane strain transferred from the actuator to the ferromagnetic
film, this is not sufficient to quantitatively analyze the indirect
magnetoelectric effect. As stressed out in the previous section, in
fact, the magnetoelastic anisotropy of the Finemet\textregistered{}
material depends directly on the Young's modulus $E$ and the Poisson's
ratio $\nu$ of the thin film. To determine them, we performed a Brillouin
light scattering (BLS) study

\subsubsection{Elastic coefficients estimation by Brillouin light scattering (BLS)}

\begin{figure}
\includegraphics[bb=0bp 0bp 379bp 275bp,clip,width=8.5cm]{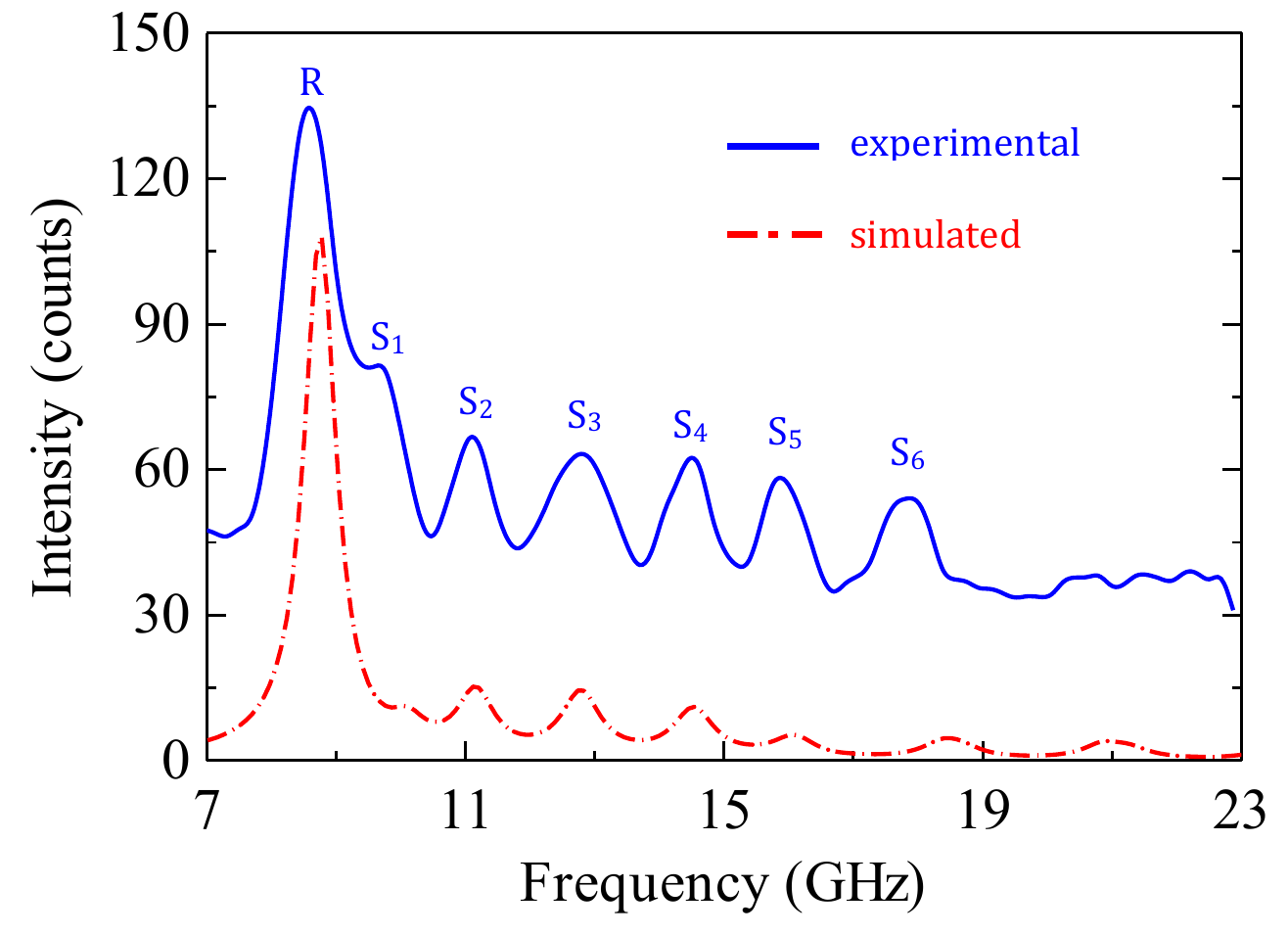}

\caption{Brillouin light scattering spectrum of the amorphous Finemet\textregistered{}
film on Kapton\textregistered{} substrate. The transferred wave-vector
is $q=2.14\times10^{5}$ cm$^{-1}$. $R$ denotes Rayleigh surface
wave while $S_{i}$ correspond to the Sesawa guided waves. The red
dashed line corresponds to the theoretical calculated spectrum while
the continuous blue line corresponds to the experimental data.}

\label{Fig_BLS} 
\end{figure}

During the last twenty years, BLS has proved to be very efficient
for achieving a complete elastic characterization of thin films and
multilayered structures \cite{Rossignol2004,Pham2013,Fillon2014}.
In a BLS experiment, a monochromatic light beam probes and reveals
acoustic phonons characteristic of the investigated medium. The power
spectrum of these phonons is mapped out from the frequency analysis
of the light scattered within a solid angle. Because of the wave vector
conservation in the phonon-photon interaction, the wavelength of the
revealed elastic waves is of the same order of magnitude as that of
light. This means that the wavelength is much larger than the inter-atomic
distances, so that the material can be described as a continuum within
an effective-medium approach. The BLS spectra were collected in air
at room temperature with typical acquisition times of a few hours.
30 mW p-polarized monochromatic light coming from a solid state laser
($\Lambda=532$ nm) was focused on the surface of the sample. We used
back-scattering confi{}guration and the scattered light was analyzed
thanks to a (3+3)-pass tandem Fabry–Pérot interferometer, so that
the value of the wave vector of the probed surface acoustic waves
is experimentally fixed to the value $q=\frac{4\pi}{\lambda}\sin\left(\xi\right)=2.14\times10^{5}$
rad.cm$^{-1}$, where $\xi=65\text{\textdegree}$ is the incidence
angle of the light beam.

For nearly opaque films with thicknesses around the acoustic wavelength
(0.3 - 0.4 $\mu$m), we can observe the surface acoustic waves with
a sagittal polarization. The Rayleigh wave ($R$), the so-called Sesawa
guided waves ($S_{1}$ to $S_{i}$) and the corresponding phase velocities
are then measured. In our case, the amorphous film can be considered
as isotropic. Thus, two isotropic elastic coefficients (Young's modulus
$E$ and Poisson's ratio $\nu$ of the thin amorphous film) influence
the Rayleigh and Sezawa modes, so that they can be evaluated by a
best fit procedure of the experimental velocities to the calculated
dispersion curves. By taking into account only the ripple mechanism
for the scattered intensity by the surface acoustic waves, the experimental
spectra are well fitted, leading to the following values : $E=145\times10^{10}$
dyn.cm$^{-2}$ ($\equiv145$ GPa) and $\nu=0.27$. The estimated Young's
modulus is close to the one of amorphous bulk FeCuNbSiB rubbans \cite{Kobelev1998}.

\subsubsection{Magnetic parameters at zero applied voltage}

\begin{figure}
\includegraphics[bb=30bp 280bp 430bp 580bp,clip,width=8.5cm]{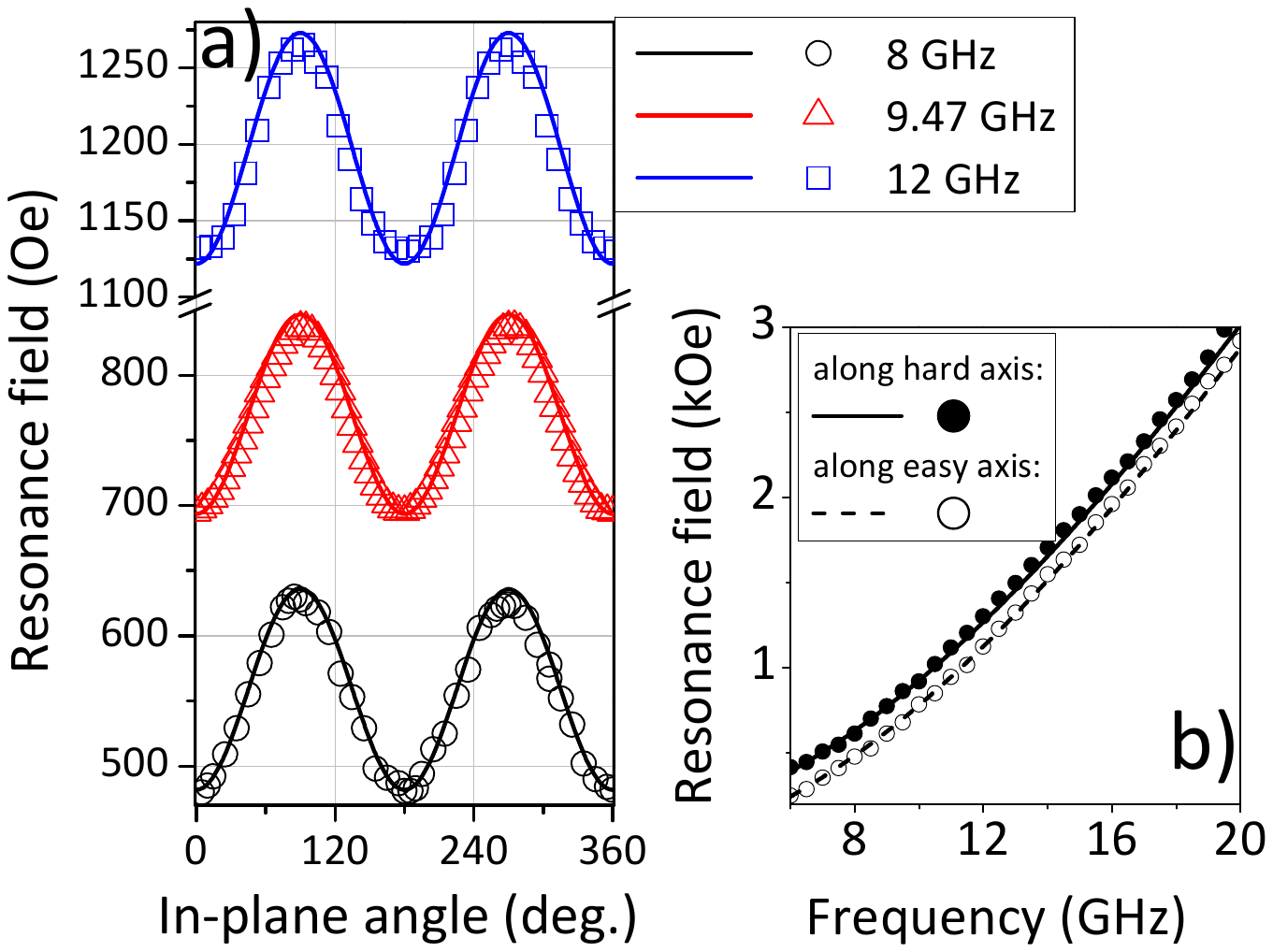} 

\caption{a) In-plane angular ($\varphi_{H}$) dependence of the resonance field
at three different driven frequencies (8, 9.47 and 12 GHz). b) Variation
of the uniform precession resonance field variation as a function
of the microwave driven frequency $f$ at zero applied voltage. Open
symbols are obtained with $\varphi_{H}=0$ (easy axis: $\vec{H}\parallel\vec{e}_{x}$)
and the filled ones are obtained with $\varphi_{H}=\pi/2$ (hard axis:
$\vec{H}\parallel\vec{e}_{y}$). The solid and dashed lines in a)
and b) are best fits to the experimental data using equation \ref{eq:Resonance_field}
with the following parameters: $\gamma=1.885\times10^{7}$ Hz.Oe$^{-1}$,
$M_{S}=965$ emu.cm$^{-3}$, $K_{u}=3.8\times10^{4}$ erg.cm$^{-3}$
and $\sigma_{xx}=\sigma_{yy}=0$.}

\label{Fig_Resonance_0V} 
\end{figure}

During this study, the magnetic properties have been probed by using
Micro-Strip FerroMagnetic Resonance (MS-FMR) which is now a common
technique to scrutinized the dynamic magnetic properties in the microwave
regime. This setup allows the determination of the resonance field
$H_{res}$ of the uniform precession mode by sweeping the applied
magnetic field in presence of a fixed pumping radio frequency fi{}eld
$\vec{h}_{rf}$ (i. e. microwave driving frequency $f$).

In order to enhance the signal to noise ratio, a weak modulation of
the static applied magnetic field (here $\sim5$ Oe at 175 Hz) is
performed. Thus, this setup gives access to the field fi{}rst derivative
of the rf absorption as a function of the applied magnetic fi{}eld.
We have first studied the magnetic properties of the Finemet\textregistered{}
thin film in zero-applied voltage. The angular ($\varphi_{H}$) dependence
of the resonance field has been studied for different microwave driving
frequency $f$. This dependence is presented in Figure \ref{Fig_Resonance_0V}a)
for three frequencies (8, 9.47 and 12 GHz). Note that the measurements
performed at 9.47 GHz have been performed by using a resonant cavity
thanks to an Electron Paramagnetic Resonance (EPR) set-up and are
clearly consistent with the MS-FMR ones. All the measurements show
that the angular dependencies are governed by a uniaxial anisotropy
having an easy axis along $x$ direction (the resonance field is indeed
minimum at $\varphi_{H}=0$). Figure \ref{Fig_Resonance_0V}b) presents
the resonance field variation as function of the microwave driving
frequency $f$ along the easy axis ($\varphi_{H}=0$, open circles)
and along the hard axis ($\varphi_{H}=\frac{\pi}{2}$, filled circles).
The continuous lines in Figure \ref{Fig_Resonance_0V} are calculated
thanks to equation \ref{eq:Resonance_field}. The in-plane strains
are equal to zero ($\varepsilon_{xx}(V=0)=\varepsilon_{yy}(V=0)=0$)
in absence of applied voltage and therefore the resonance field depends
only on $\gamma$, $M_{s}$ and $K_{u}$ \cite{Zighem2010}. The best
fits to experimental data are obtained using the following parameters:
$\gamma=1.885\times10^{7}$ s$^{-1}$.Oe$^{-1}$, $M_{S}=965$ emu.cm$^{-3}$
(i. e. $4\pi M_{s}\sim12100$ G) and $K_{u}=3.8\times10^{4}$ erg.cm$^{-3}$(i.
e. $H_{u}\sim80$ Oe). The obtained Landé factor ($g\sim2.1)$ value
is the typical for metallic ferromagnets and the $M_{s}$ value is
in a good agreement with previously determined values for similar
films. However, due to the amorphous structure of the material, the
in-plane magnetic anisotropy in equivalent Finemet\textregistered{}
film deposited onto rigid substrate (Si) is generally weaker (a few
Oe) than the $\sim80$ Oe found here. As previously suggested \cite{ZhangJAP_2013,ZhangAPL_2013,Gueye2014,Gueye2014_BIS},
the origin of this uniaxial anisotropy is certainly due to a non zero
magnetoelastic anisotropy at zero-applied voltage. This {}``initial''
anisotropy can be due to a slight curvature along a given direction
taking place during the elaboration process which could lead to a
magnetoelastic anisotropy at zero-applied voltage. Thereafter, this
residual anisotropy will be modeled as an \textit{ad hoc} uniaxial
one.

\section{Magnetic anisotropy strains effect }

\subsection{Measured in-plane strains }

\begin{figure*}
\includegraphics[bb=30bp 400bp 760bp 595bp,clip,width=15cm]{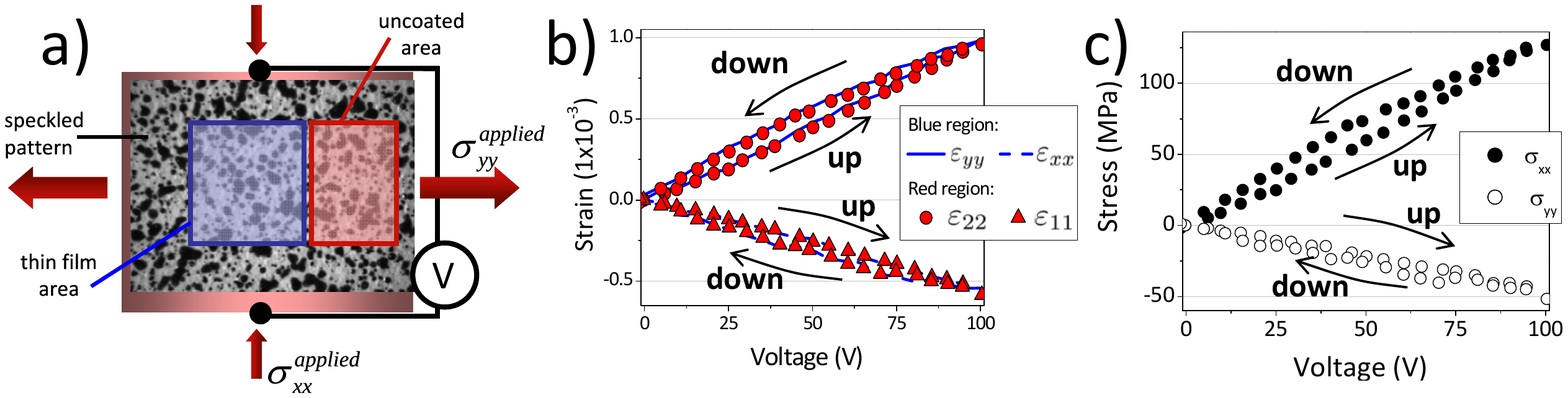}
\caption{a) Top view sketch of the system showing the spray-painted speckle
pattern in order to generate a contrast at the surface of the film.
The blue and red areas correspond to the top surface of the film region
and to an uncoated area of the actuator, respectively. b) Extracted
mean in-plane strains from the blue (film surface) and red (uncoated
actuator surface) areas by performing DIC calculations. Continuous
lines correspond to $\varepsilon_{xx}$ and $\varepsilon_{yy}$ while
and symbols correspond to $\varepsilon_{11}$ and $\varepsilon_{22}$.
c) Calculated mean in-plane stresses $\sigma_{xx}$ and $\sigma_{yy}$
using equations \ref{eq:Hook_law_1} and \ref{eq:Hook_law_2} with
$E=145\times10^{10}$ dyn.cm$^{-2}$ and $\nu=0.27$. In b) and c),
the applied external voltage was swept from 0 V to 100 V and back
to 0 V with steps of around 5 V. The first measurement at 0 V has
been performed after {}``saturating'' the actuator at 100 V to avoid
training effect of the actuator polarization.}

\label{Fig_Strain_Stress} 
\end{figure*}

The piezoelectric actuator is characterized by a main axis direction
(direction 2 in Figure \ref{Fig_Sketch_Actuator}). The film/substrate
system has been glued in order to have the main deformation axis of
the actuator perpendicular to the easy axis, so in this condition,
the direction $1$ (resp. $2$) of the actuator refers to $x$ (resp.
$y$) direction of the thin film. Digital image correlation (DIC)
method enables accurate measurements of changes in digital images
\cite{Haddadi2012,Djaziri2011}. This method uses tracking and image
registration to make full-field non-contact measurements of displacements
and strains in a wide variety of engineering applications (mechanics
of materials, micro- and nano-technology, ...). Here, two CCD cameras
mounted on a tripod have been positioned vertically in top of the
film/substrate/actuator system, the field of view is fixed to approximately
$2\times2$ cm$^{2}$. Given the $2448\times2050$ number of pixels
of the cameras, the area per pixel is about $12.5$ nm$^{2}$. The
respective positions of the cameras has been calibrated by using a
10 mm$\times$8 mm calibration pattern. Figure \ref{Fig_Strain_Stress}a)
presents a part speckle pattern which has been generated at the {}``uniform''
top surface of the heterostructure by using a spray paint in order
to generate a contrast which will serve to calculate the deformation.
Indeed, a first image (reference image) is taken at zero applied voltage;
then, a sequence of images are taken at different applied voltages
and are compared to the reference. The field strain at the surface
of the system has been extracted for each applied voltage by performing
DIC calculations which are performed by using the reference image
and the different images coming from the sequence. The DIC calculations
have been performed by using ARAMIS which is a commercially available
software package \cite{Aramis}. From the fields strain, the mean
in-plane strains are extracted as function of the applied voltage.
Note that shear strains can also be extracted because of the use of
two cameras in the present setup. However, the shear strains values
are found to be negligible in this study and will be neglected thereafter.

Images were collected by performing applied voltage loops (from 0V
to 100V and back to 0V) with a frame rate of about 0.1 FPS; the step
of applied voltage was fixed to around 5 V. Furthermore, prior to
the measurements, several images have been taken in absence of voltage
in order to estimate the DIC setup statistical errors, estimated to
be $\sim5\times10^{-6}$. Moreover, different images have been taken
as a function of time at 0 V after saturating the actuator at 100
V. After approximately 5 hours, a difference of about $\left|4\times10^{-5}\right|$
in the in-plane strains values is found; this value rises to $\left|1\times10^{-4}\right|$
after several days (which is relatively high). This difference is
due to the training effect of the polarization. In addition, ARAMIS
has been used to calculate the DIC in two different $3\times3$ mm$^{2}$
regions: an uncoated area of the actuator and an area located at the
top of the thin film.

Figures \ref{Fig_Strain_Stress}b) presents the extracted mean in-plane
strains $\varepsilon_{11}$ and $\varepsilon_{22}$. Indeed, similar
quasi-homogeneous strain fields as function of the applied voltage
have been calculated from the two regions. Thus a $100\%$ in-plane
strain transmission in between the piezoelectric actuator and the
film is observed, it can be conclude that $\varepsilon_{11}=\varepsilon_{xx}$
and $\varepsilon_{22}=\varepsilon_{yy}$. Non linear and hysteretic
variations are observed for both $\varepsilon_{11}$ and $\varepsilon_{22}$
which is certainly due to the intrinsic properties of the ferroelectric
material used in the fabrication of the actuator \cite{Zighem_JAP2013}.
One can note that $\varepsilon_{22}$ is found to be positive and
$\varepsilon_{11}$ is found to be negative in the voltage range {[}0-100
V{]}. Moreover, it is interesting to note that a linear variation
of $\varepsilon_{22}$ as a function of $\varepsilon_{11}$ is found
$\varepsilon_{22}\simeq-1.7\varepsilon_{11}$. The maximum achieved
values of $\varepsilon_{22}$ and $\varepsilon_{11}$ ($\sim1\times10^{-3}$
$\sim-0.5\times10^{-3}$ at 100 V, respectively) show that the film
is not deteriorated by the plasticity regime because it is obtained
for higher values: this is experimentally confirmed by the excellent
reproducibility of the experiments (even after several days). In this
condition and because of the amorphous structure of the magnetic film,
the in-plane stresses can be calculated using equations \ref{eq:Hook_law_1}
and \ref{eq:Hook_law_2} with $E$ and $\nu$ values previously determined
by BLS. Obviously, $\sigma_{11}$ and $\sigma_{22}$ also present
non linear and hysteretic variations as function of the applied voltage.
This representation of the in-plane stresses could be useful, especially
in the case of uniaxial in-plane stress which is not the case here.
Indeed, for $\sigma_{22}$ , a similar linear variation is found with
$\sigma_{22}\simeq-3.6\sigma_{11}$.

From the voltage dependence of $\varepsilon_{11}$ and $\varepsilon_{22}$
(or $\sigma_{11}$. and $\sigma_{22}$), the indirect magnetoelectric
effect can be quantitatively studied in the the system by probing
resonance field uniform mode.

\begin{figure}
\includegraphics[bb=25bp 20bp 490bp 590bp,clip,width=8.5cm]{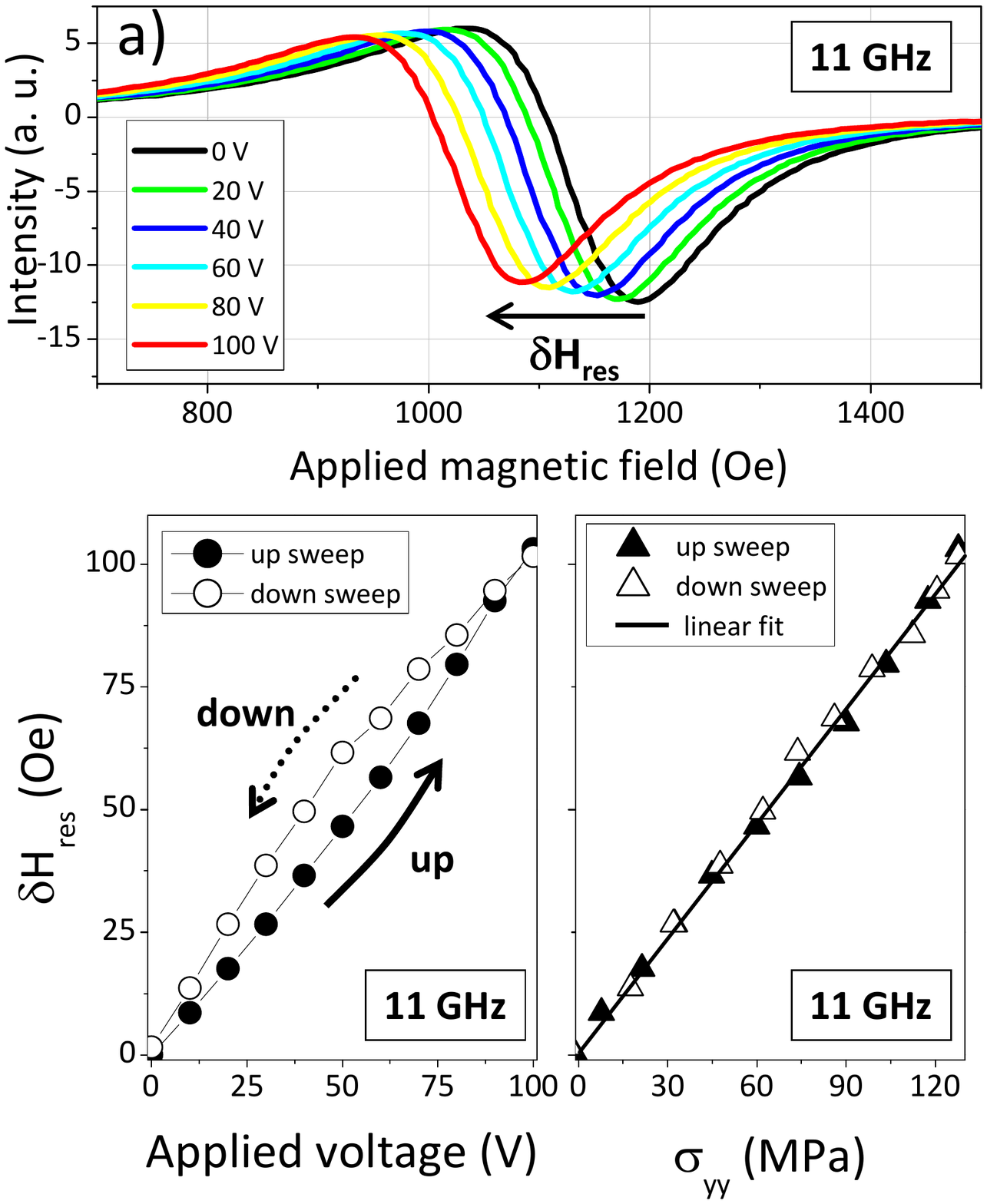}\caption{a) Experimental spectra recorded at 11 GHz with $\varphi_{H}=\frac{\pi}{2}$
at different applied voltages. (b) Resonance field variations (at
11 GHz and $\varphi_{H}=0$) as a function of the applied voltage
(b) and of the applied stress (c). The open symbols correspond to
upsweep (0 to 100 V) while filled symbols correspond to downsweep
(100 to 0 V). Note that $\delta H_{res}$ hysteresis disappears when
the data are plotted as function of $\sigma_{yy}$ (or $\sigma_{xx}$).
The solid lines in Figure 7b are guided for the eyes while the solid
line in Figure 7c corresponds to a linear fit.}

\label{Fig_FMR_Spectra} 
\end{figure}

\begin{table}[!t]
\begin{singlespace}
\noindent %
\begin{tabular}{>{\raggedright}m{1.4cm}>{\centering}p{1.5cm}>{\centering}p{1.4cm}>{\centering}p{1.2cm}>{\centering}p{1.4cm}}
 & \multicolumn{2}{c}{$\alpha_{me}$ (V.cm$^{-1}$.Oe$^{-1}$)} & \multicolumn{2}{c}{$\alpha_{ms}^{\sigma_{yy}}$ (Oe.MPa$^{-1}$)}\tabularnewline
\hline 
\hline 
$f$ (GHz)  & \multicolumn{4}{c}{$\varphi_{H}$}\tabularnewline
 & $0$  & $\frac{\pi}{2}$  & $0$  & $\frac{\pi}{2}$\tabularnewline
\hline 
\hline 
8 & -1.62  & 1.4  & -0.67  & 0.77\tabularnewline
10 & -1.62  & 1.39  & -0.62  & 0.77\tabularnewline
11 & -1.76  & 1.41  & -0.63  & 0.77\tabularnewline
12 & -1.68  & 1.41  & -0.64  & 0.76\tabularnewline
13  & -1.7  & 1.4  & -0.58  & 0.75\tabularnewline
14  & -1.76  & 1.41  & -0.62  & 0.76\tabularnewline
15  & -1.85  & 1.44  & -0.58  & 0.75\tabularnewline
16  & -1.81  & 1.76  & -0.59  & 0.62\tabularnewline
18  & -1.81  & 1.7  & -0.6  & 0.64\tabularnewline
20  & -1.831  & 1.37  & -0.6  & 0.79\tabularnewline
\hline 
\hline 
 &  &  &  & \tabularnewline
\end{tabular}
\end{singlespace}

\caption{$\alpha_{me}$ and $\alpha_{ms}^{\sigma_{yy}}$ as function of the
microwave driving frequency $f$ extracted along the $x$ and $y$
direction.}

\label{Table_1} 
\end{table}

\begin{figure*}
\includegraphics[bb=20bp 190bp 770bp 570bp,clip,width=14cm]{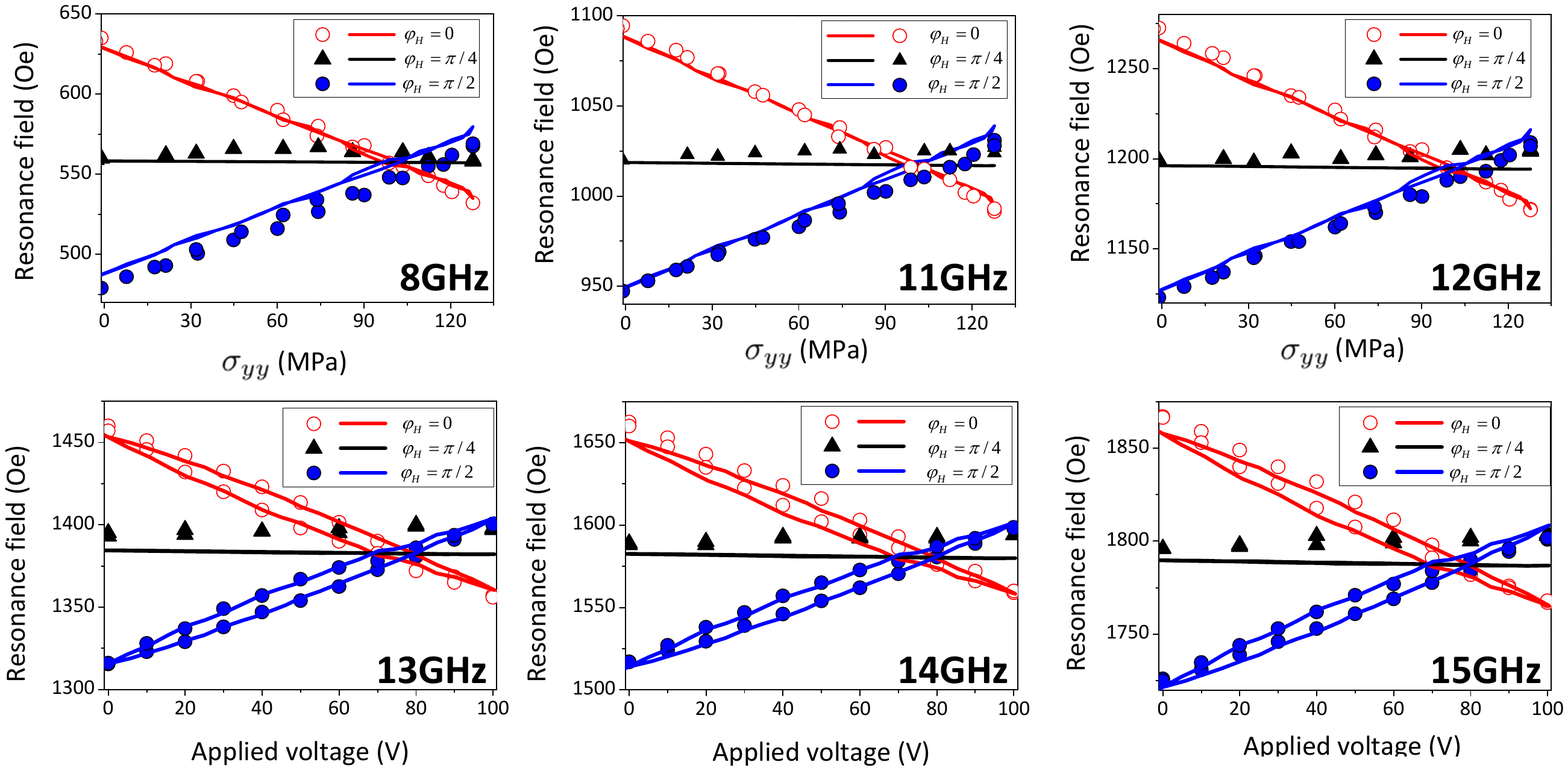}\caption{Resonance field variation as a function of the induced in-plane stress
$\sigma_{22}$ (top graphs) and the applied voltage (down graphs)
measured at different microwave driven frequencies (8-15 GHz) and
in-plane angular angles ($\varphi_{H}$=0, $\frac{\pi}{2}$ and $\frac{\pi}{4}$).
Note that the observed hysteretic and non linear variations (see Figure
\ref{Fig_FMR_Spectra} disappears). Open circles are obtained with
$\varphi_{H}=0$ (initial easy axis: $\vec{H}\parallel\hat{x}$),
the filled circles are obtained with $\varphi_{H}=\frac{\pi}{2}$
(hard axis: $\vec{H}\parallel\hat{y}$) and the open triangles are
obtained with $\varphi_{H}=\frac{\pi}{4}$. The different lines are
best fits to the experimental data using equation \ref{eq:Resonance_field}
with the following parameters: $\gamma=1.885\times10^{7}$ s$^{-1}$.Oe$^{-1}$,
$M_{S}=965$ emu.cm$^{-3}$, $K_{u}=3.8\times10^{4}$ erg.cm$^{-3}$,
$E=145$ GPa and $\nu=0.27$ and $\lambda=16\times10^{-6}$.}

\label{Fig_Resonance_Voltage} 
\end{figure*}

\subsection{Strain induced anisotropy and magnetoelastic behavior}

\begin{figure}[!h]
\includegraphics[bb=30bp 0bp 440bp 580bp,clip,width=8.5cm]{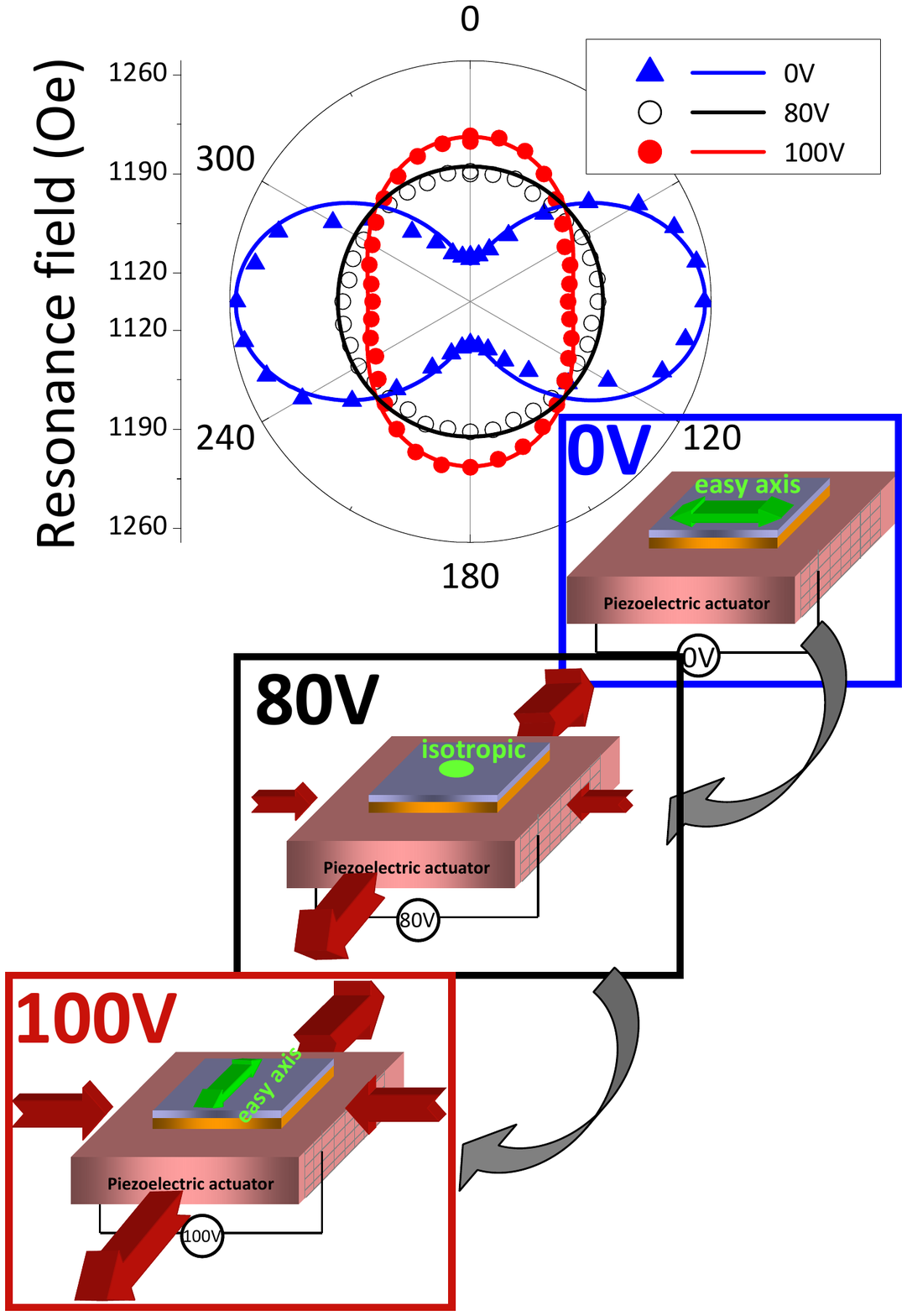}\caption{In-plane angular ($\varphi_{H}$) dependence of the resonance field
measured at 12 GHz for three different applied voltages: 0, 80 and
100 V. The symbols are experimental data (triangles: 0 V, open circles:
80 V and filled circles: 100 V) while the solid lines are calculated
thanks to equation \ref{eq:Resonance_field} with the following parameters:
$\gamma=1.885\times10^{7}$ s$^{-1}$.Oe$^{-1}$, $M_{S}=965$ emu.cm$^{-3}$,
$K_{u}=3.8\times10^{4}$ erg.cm$^{-3}$, $E=145\times10^{10}$ dyn.cm$^{-2}$
and $\nu=0.27$ and $\lambda=16\times10^{-6}$. The Sketches correspond
to 3D view of the heterostructure showing the voltage-switch of the
easy axis from $x$ to $y$ direction.}

\label{Fig_Resonance_Voltage_Angular} 
\end{figure}

Figures \ref{Fig_FMR_Spectra}a) shows typical MS-FMR experimental
spectra recorded at 11 GHz with an applied magnetic field along $y$
direction (initial hard axis: $\varphi_{H}=\frac{\pi}{2}$) at different
applied voltages (0, 20, 40, 60, 80 and 100 V). A shift of the resonance
field $\delta H_{res}$ of around 100 Oe between spectra recorded
at 0 and 100 V is clearly obvious. The corresponding variation of
$\delta H_{res}$ (defined as $\delta H_{res}=H_{res}(0)-H_{res}(V)$)
as a function of the applied voltage is reported in Figure \ref{Fig_FMR_Spectra}:
a non linear and hysteretic variation is observed. Here, filled circles
represent the upsweep (0 to 100 V) and open circles the downsweep
of the applied voltage (100 to 0 V). The spectra presented in Figure
\ref{Fig_FMR_Spectra}a) show that the resonance field decreases when
increasing the applied voltage. Since $\varepsilon_{yy}>0$ and $\varepsilon_{xx}<0$,
it can be conclude that the magnetostriction coefficient at saturation
$\lambda$ of the thin film is positive. Indeed, a negative magnetostriction
coefficient would have led to an increase of $H_{res}$, as it is
the case in Ni polycrystalline thin film \cite{Zighem_JAP2013,BrandlmaierNewJ.Phys_2009}.

Furthermore, a linear variation of $\delta H_{res}$ appears if the
voltage-stress dependence (see Figure \ref{Fig_Strain_Stress}) is
used. Figure \ref{Fig_FMR_Spectra}c) illustrates such a behavior
where the continuous line is the linear fit of the slope which is
found around $\alpha_{ms}^{\sigma_{yy}}=0.76$ Oe.MPa$^{-1}$. This
feature indicates that the non linear and the hysteretic variations
of $\delta H_{res}$ as a function of $V$ is not related to the magnetoelastic
anisotropy; it is completely due to the intrinsic properties of the
ferroelectric material used for the actuator fabrication. In addition,
in first approximation, if the non linear and the hysteretic behavior
of $\delta H_{res}$ as a function of $V$ is neglected and adjusted
by a linear fit, an effective magnetoelectric coupling ($\alpha_{me}$
in V.cm$^{-1}$.Oe$^{-1}$ ) can be estimated (by considering the
static electric field inside the actuator. Table \ref{Table_1} presents
the different values of $\alpha_{me}$ and $\alpha_{ms}^{\sigma_{yy}}$
as function of the microwave driving frequency measured along the
initial easy and hard axes ($\varphi_{H}=0$ and $\frac{\pi}{2}$).
The first observation is that the voltage-induced magnetoelastic effect
is frequency independent for this system. However, a weak difference
of $\alpha_{me}$ extracted from the measurements performed at $\varphi_{H}=0$
and $\varphi_{H}=\frac{\pi}{2}$ is found. Such effect is most probably
due to the misalignment (of a few degrees) of the $x$ direction and
the direction $1$ of the actuator which may occurred when gluing
of the film/substrate system onto the piezoelectric actuator and is
predicted by the equation \ref{eq:Resonance_field}.

Figure \ref{Fig_Resonance_Voltage} presents variations of $H_{res}$
as a function of the applied voltage-induced stress (three top graphs)
and as a function of the applied voltage (three down graphs) for different
$\varphi_{H}$ angles ($0$, $\frac{\pi}{4}$ and $\frac{\pi}{2}$)
at various microwave driving frequencies. In first approximation,
the induced magnetoelastic anisotropy can be viewed as a uniaxial
magnetoelastic anisotropy field $\vec{H}_{me}$ along $y$ direction,
which is thus perpendicular to the {}``initial'' uniaxial anisotropy
field $\vec{H}_{u}$. In absence of applied voltage, the resonance
field along $x$ direction ($\varphi_{H}=0$) is smaller than the
one measured along $y$ direction ($\varphi_{H}=\frac{\pi}{2}$),
the difference of these two last resonance fields is roughly equal
to $\left\Vert 2\vec{H}_{u}\right\Vert $. When increasing the applied
voltage (or induced-stress), the $x$ direction will be less easy
which leads to an increase (resp. decrease) of the resonance field
along $x$ (resp. along $y$) direction because of the competition
between $\vec{H}_{me}$ and $\vec{H}_{u}$. At 80 V ($\sigma_{11}\sim-30$
MPa and $\sigma_{22}\sim105$ MPa), the resonance fields are equal
for the three studied angles which means that $\vec{H}_{u}$ is totally
compensated by $\vec{H}_{me}$. A quantitative analysis of the voltage-induced
variation of the resonance field has been performed by using equation
\ref{eq:Resonance_field}. By introducing the stress-voltage dependence,
the only undetermined parameter is the saturation magnetostriction
coefficient $\lambda$; the best fits of all the experiments gives
$\lambda=16\times10^{-6}$, which is slightly lower with the bulk
material. A good agreement is found between the experimental variations
and the calculated ones for the different frequencies (see \ref{Fig_Resonance_Voltage}).
Note that the non linear and the hysteretic variations of $H_{res}$
as a function of $V$ are well reproduced. In addition, at $\varphi_{H}=\frac{\pi}{4}$,
an experimentally confirmed almost constant variation as predicted
by equation \ref{eq:Resonance_field}.

Figure \ref{Fig_Resonance_Voltage_Angular} presents angular variations
of the resonance field at different applied voltage: 0, 80 and 100V
which correspond to $\sigma_{11}\sim0$, -30 and -45 MPa and $\sigma_{22}\sim0$,
105 and 130 MPa, respectively. At zero-voltage, the angular variation
of the resonance field found in Figure \ref{Fig_Resonance_0V}a) is
retrieved, the horizontal peanut shape represented by blue triangles
is characteristic of a uniaxial anisotropy along $x$ direction ($\varphi_{H}=0$).
The 80 V value was chosen because of the {}``exact'' compensation
of this uniaxial anisotropy by the induced magnetoelastic one. At
this voltage, the film is in-plane isotropic in the magnetic point
of view. It is experimentally confirmed by the circle shape formed
by the open circles. Finally, at 100 V, a uniaxial anisotropy characterized
by a vertical peanut shape is observed along $y$ axis ($\varphi_{H}=\frac{\pi}{2}$).
Thus, a voltage-switch of the effective easy axis from $x$ direction
(at 0 V) to $y$ direction (at 100 V) has been performed. The sketches
of Figure \ref{Fig_Resonance_0V}b qualitatively present such effect.
Finally, the solid lines of Figure \ref{Fig_Resonance_0V}a are calculated
thanks to equation \ref{eq:Resonance_field} with the parameters previously
determined.

\section{Conclusions}
\begin{itemize}
\item Magnetic anisotropy in Finemet\textregistered{} thin films deposited
on Kapton\textregistered{} substrate has been studied by Micro-Strip
FerroMagnetic Resonance (MS-FMR). 
\item We have shown that the flexibility of Kapton\textregistered{} substrate
allowed tailoring the magnetic anisotropy of the film by applying
small voltage-induced strains. 
\item The flexibility of the Kapton\textregistered{} substrate induced an
initial uniaxial anisotropy that is generally not found in ferromagnetic
films whose thickness is a few hundred nanometers. 
\item The knowledge of the applied elastic strains \textit{versus} applied
voltage measured by Digital Image Correlation and Finemet\textregistered{}
film elastic constants measured by Brillouin light Scattering allowed
estimating the effective magnetostriction coefficient of the film.\end{itemize}
\begin{acknowledgments}
The authors gratefully acknowledge the CNRS for his financial support
through the {}``PEPS INSIS'' program (FERROFLEX project) and the
Renatech network supporting the IEF clean room facilities. This work
has been also partially supported by the French Research Agency (ANR)
in the frame of the project ANR 2010 JCJC 090601 entitled {}``SpinStress''
and by the Université Paris 13 through a {}``Bonus Qualité Recherche''
project. The authors thank Pr. Dr. Philippe Djemia for discussion
concerning Brillouin Ligth Scattering measurements and analysis. The
authors are also grateful to Dr. Brigitte Leridon (LPEM-ESPCI, ParisTech)
for putting at their disposal the experimental EPR setup.\end{acknowledgments}


\begin{thebibliography}{References}
\bibitem{Nan2011_Adv_Mat} Jing Ma , Jiamian Hu , Zheng Li and Ce-Wen
Nan, Adv. Mater. \textbf{23}, 1062 (2011)

\bibitem{Ramesh2010_Adv_Mat} Carlos A. F. Vaz , Jason Hoffman , Charles
H. Ahn and Ramamoorthy Ramesh, Adv. Mater., \textbf{22}, 2900 (2010)

\bibitem{Martins2013_Adv_Func_Mater} Pedro Martins and Senentxu Lanceros-Méndez,
Adv. Funct. Mater., \textbf{23}, 3371 (2013)

\bibitem{Ma2011} J. Ma, J. Hu, Z. Li, C.-W. Nan, Adv. Mater. \textbf{23},
1062-1087 (2011)

\bibitem{Lahtinen2011} T. H. E. Lahtinen, J. O. Tuomi, S. van Dijken,
Adv. Mater. \textbf{23}, 3187-3191 (2011)

\bibitem{Roy2011}K. Roy, S. Bandyopadhyay, J. Atulasimha, Appl. Phys.
Letters \textbf{99}, 063108 (2011)

\bibitem{Pettiford2008} C. Pettiford, J. Lou, L. Russell, and N.
X. Sun, App. Phys. Lett. \textbf{92}, 122506 (2008)

\bibitem{Brandlmaier2008_PRB} C. Bihler, M. Althammer, A. Brandlmaier,
S. Geprägs, M. Weiler, M. Opel, W. Schoch, W. Limmer, R. Gross, M.
S. Brandt and S. T. B. Goennenwein, Phys. Rev. B \textbf{78}, 045203
(2008)

\bibitem{Zighem_JAP2013} F. Zighem, D. Faurie, S. Mercone, M. Belmeguenai
and H. Haddadi, J. App. Phys. \textbf{114}, 073902 (2013)

\bibitem{Tiercelin2011} N. Tiercelin, Y. Dusch, A. Klimov, S. Giordano,
V. Preobrazhensky and P. Pernod, App. Phys. Lett., \textbf{99}, 192507
(2011)

\bibitem{Brandlmaier2008_PRB_Bis} A. Brandlmaier, S. Geprägs, M.
Weiler, A. Boger, M. Opel, H. Huebl, C. Bihler, M. S. Brandt, B. Botters,
D. Grundler, R. Gross and S. T. B. Goennenwein, Phys. Rev. B \textbf{77},
104445 (2008)

\bibitem{Barrault2012}C. Barraud, C. Deranlot, P. Seneor, R. Mattana,
B. Dlubak, S. Fusil, K. Bouzehouane, D. Deneuve, F. Petroff, and A.
Fert, Appl. Phys. Lett. \textbf{96}, 072502 (2010)

\bibitem{Bedoya-Pinto2014}A. Bedoya-Pinto, M. Donolato, M. Gobbi,
L. E. Hueso, Paolo Vavassori, App. Phys. Lett. \textbf{104}, 062412
(2014)

\bibitem{ZhangJAP2013_Bis} Z. Zuo,1,2 Q.Zhan, G. Dai, B. Chen, X.
Zhang, H. Yang, Y. Liu and R.-W. Li, J. Appl. Phys., \textbf{113},
17C705 (2013)

\bibitem{Smit_Beljers_1955} J. Smit, H.G. Beljers, Philips Res. Rep.
\textbf{10}, 113 (1955)

\bibitem{Suhl_1955} H. Suhl, Phys. Rev. \textbf{97}, 555 (1955)

\bibitem{Rossignol2004} C. Rossignol, B. Perrin, B. Bonello, P. Djemia,
P. Moch, H. Hurdequint, Phys. Rev. B \textbf{70}, 094102 (2004)

\bibitem{Fillon2014}A. Fillon, C. Jaouen, A. Michel, G. Abadias,
C. Tromas, L. Belliard, B. Perrin, P. Djemia, Phys. Rev. B \textbf{88},
174104 (2014)

\bibitem{Pham2013} T. Pham, D. Faurie, P. Djemia, L. Belliard, E.
Le Bourhis, P. Goudeau, F. Paumier, Appl. Phys. Letter \textbf{103},
041601 (2013)

\bibitem{belmeguenai2009} M. Belmeguenai, F. Zighem, Y. Roussigné,
S-M. Chérif, P. Moch, K. Westerholt, G. Woltersdorf and G. Bayreuther,
Phys. Rev. B \textbf{79}, 024419 (2009)

\bibitem{belmeguenai2013} M. Belmeguenai, H. Tuzcuoglu, M. S. Gabor,
T. Petrisor jr, C. Tiusan, D. Berling, F. Zighem, T. Chauveau, S.
M. Chérif, P. Moch, Phys. Rev. B \textbf{87}, 184431 (2013)

\bibitem{Kobelev1998}N.P Kobelev, Y. M. Soifer, Nanostructured Materials
\textbf{10}, 449-456 (1998)

\bibitem{Yoshizawa}Y. Yoshizawa, S. Oguma, and K. Yamauchi, J. Appl.
Phys. \textbf{64}, 6044 (1988)

\bibitem{Herzer1993}G. Herzer, Physica Scripta \textbf{T49}, 307
(1993)

\bibitem{Zighem2010} F. Zighem, Y. Roussigné, S. M. Chérif, P. Moch,
J. Ben Youssef, and F. Paumier, J. Phys.: Condens. Matter. \textbf{22},
406001 (2010).

\bibitem{ZhangJAP_2013} X. Zhang, Q. Zhan, G. Dai, Y. Liu, Z. Zuo,
H. Yang, B. Chen, and R.-W. Li, J. Appl. Phys., \textbf{113}, 17A901
(2013).

\bibitem{ZhangAPL_2013} X. Zhang, Q. Zhan, G. Dai, Y. Liu, Z. Zuo,
H. Yang, B. Chen and R.-W. Li, App. Phys. Lett., \textbf{102}, 022412
(2013)

\bibitem{Gueye2014} M. Gueye, F. Zighem, D. Faurie, M. Belmeguenai
and S. Mercone, App. Phys. Lett., \textbf{105}, 052411 (2014)

\bibitem{Gueye2014_BIS} M. Gueye, B. M. Wague, F. Zighem, M. Belmeguenai,
M. S. Gabor, T. Petrisor Jr, C. Tiusan, S. Mercone and D. Faurie,
App. Phys. Lett., \textbf{105}, 062409 (2014)

\bibitem{Haddadi2012} H. Haddadi and S. Belhabib, Int. J. of Mech.
Sc. \textbf{62}, 47 (2012)

\bibitem{Djaziri2011} S. Djaziri, P. O. Renault, F. Hild, E. Le Bourhis,
P. Goudeau, D. Thiaudière, D. Faurie, J. Appl. Cryst. \textbf{44},
1071 (2011)

\bibitem{Aramis}  http://www.gom.com/

\bibitem{BrandlmaierNewJ.Phys_2009} M. Weiler, A. Brandlmaier, S.
Gepr€ags, M. Althammer, M. Opel, C. Bihler, H. Huebl, M. S. Brandt,
R. Gross, and S. T. B. Goennenwein, New J. Phys., \textbf{11}, 013021
(2009).\end{thebibliography}
\end{document}